\documentclass[conference]{IEEEtran}
\IEEEoverridecommandlockouts

\usepackage{cite}
\usepackage{amsmath,amssymb,amsfonts}
\usepackage{algorithmic}
\usepackage{graphicx}
\usepackage{textcomp}
\usepackage{xcolor}
\def\BibTeX{{\rm B\kern-.05em{\sc i\kern-.025em b}\kern-.08em
    T\kern-.1667em\lower.7ex\hbox{E}\kern-.125emX}}
\usepackage{tikz}  
\usetikzlibrary{decorations.pathreplacing,calligraphy}
\usepackage[framemethod=tikz]{mdframed}
\usetikzlibrary{shapes.geometric, arrows, arrows.meta}
\usepackage{graphicx}
\usepackage{float}
\usepackage{booktabs}
\usepackage{multirow}
\usepackage{dblfloatfix}
\usepackage{siunitx}
\DeclareSIUnit[quantity-product = ]\percent{\char`\%}
\usepackage{comment}
\usetikzlibrary{decorations.text}
\usetikzlibrary{shapes.geometric, arrows}
\usetikzlibrary{positioning}
\usepackage{bm}

\usepackage{pifont} 
\makeatletter
\newcommand*\bigcdot{\mathpalette\bigcdot@{.5}}
\newcommand*\bigcdot@[2]{\mathbin{\vcenter{\hbox{\scalebox{#2}{$\m@th#1\bullet$}}}}}
\makeatother
\usepackage{url}
\newcommand{\norm}[1]{\left\lVert#1\right\rVert}
\newcommand{\audiopure}{Audio\-Pure }
\begin{document}

\title{Detecting and Defending Against Adversarial Attacks on Automatic Speech Recognition via Diffusion Models
}

\author{\IEEEauthorblockN{Nikolai L. Kühne\IEEEauthorrefmark{1}$^\circ$, 
Astrid H. F. Kitchen\IEEEauthorrefmark{1}$^\circ$, 
Marie S. Jensen\IEEEauthorrefmark{1}$^\circ$, 
Mikkel S. L. Brøndt\IEEEauthorrefmark{1}$^\circ$,
Martin Gonzalez\IEEEauthorrefmark{2},\\
Christophe Biscio\IEEEauthorrefmark{1}, and
Zheng-Hua Tan\IEEEauthorrefmark{1}
\thanks{$^\circ$ These authors contributed equally to this work.}
}
\IEEEauthorblockA{
\IEEEauthorrefmark{1}Aalborg University, Denmark\\ 
Email: nlk@es.aau.dk, \{akitch20, marije19, mbrand20\}@student.aau.dk, christophe@math.aau.dk, zt@es.aau.dk}
\IEEEauthorblockA{
\IEEEauthorrefmark{2}IRT SystemX, France\\ 
Email: martin.gonzalez@irt-systemx.fr}
}

\maketitle

\begin{abstract}
Automatic speech recognition (ASR) systems are known to be vulnerable to adversarial attacks. 
This paper addresses detection and defence against targeted white-box attacks on speech signals for ASR systems. 
While existing work has utilised diffusion models (DMs) to purify adversarial examples, achieving state-of-the-art results in keyword spotting tasks, their effectiveness for more complex tasks such as sentence-level ASR remains unexplored. Additionally, the impact of the number of forward diffusion steps on performance is not well understood.
In this paper, we systematically investigate the use of DMs for defending against adversarial attacks on sentences and examine the effect of varying forward diffusion steps. 
Through comprehensive experiments on the Mozilla Common Voice dataset, we demonstrate that two forward diffusion steps can completely defend against adversarial attacks on sentences. 
Moreover, we introduce a novel, training-free approach for detecting adversarial attacks by leveraging a pre-trained DM. Our experimental results show that this method can detect adversarial attacks with high accuracy. 

\end{abstract}

\begin{IEEEkeywords}
Automatic speech recognition, adversarial attacks, diffusion models.
\end{IEEEkeywords}

\section{Introduction}

Automatic speech recognition (ASR) systems, like other machine learning models, are vulnerable to adversarial attacks. These attacks involve adding subtle, optimised perturbations to the original input to deceive the ASR system, while remaining imperceptible to humans \cite{DBLP:journals/corr/SzegedyZSBEGF13,43405, papernot2016limitations, cisse2017houdini, yuan2018commandersong, qin2019imperceptible, taori2019targeted, du2020sirenattack, abdullah2021hear, chen2021real}.
The consequences of such attacks are severe, potentially leading to compromised security systems, unauthorised purchases, and theft of sensitive information stored on devices controlled by voice assistants.
Given the widespread use of ASR systems in mobile phones, smart devices, and vehicles, the need to detect and defend against adversarial attacks has become increasingly urgent \cite{herff2016automatic}.

Existing defensive methods, including input transfor\-mation-based defences and distillation \cite{yang2018characterizing,rajaratnam2018speech, distillation2016, chen2022towards}, reduce adversarial attack success rate but are less effective against strong (e.g., white-box (WB)) attacks \cite{carlini2018audio,carlini2017towards} and ineffective against adaptive attacks \cite{yang2018characterizing}.  
Adversarial training \cite{madry2017towards}, considered the most effective defence according to \cite{wu2023defending}, still leaves models vulnerable to attacks not encountered during training.

In this work, we address key issues by leveraging a pre-trained diffusion model (DM) \cite{sohl2015deep, ho2020denoising, nichol2021improved} to detect and defend against targeted WB attacks on ASR systems. Targeted attacks aim to trick the ASR system into misclassifying inputs as a specific, incorrect class, with WB attacks having full access to the parameters of the target model.

By exploring the inherent denoising property of denoising diffusion probabilistic models (DDPMs) \cite{ho2020denoising}, an adversarial purification-based method for audio was proposed in \cite{wu2023defending} based on the pre-trained DDPM from \cite{kong2021diffwave}.
This method proved to be the most effective against WB attacks on keywords compared to other cutting-edge defence methods. 
However, these methods have not been tested on sentences, which ASR systems typically process.
This gap motivates our approach: applying diffusion models to sentences and exploring the impact of varying forward diffusion steps. Each step adds noise to override the adversarial perturbations, with the reverse diffusion process generating purified data from the noisy inputs, thereby enhancing the robustness of  ASR systems against adversarial attacks.


\begin{figure*}[thb]
    \centering
    \input{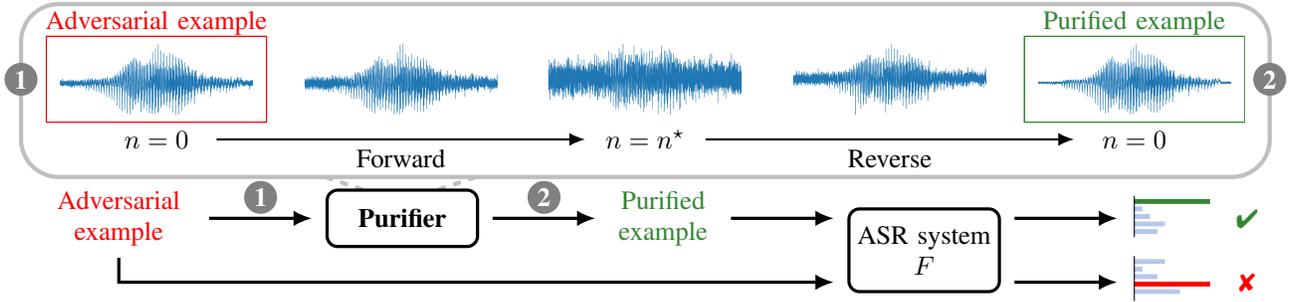}
    \caption{The architecture of the whole speech system protected by the \textbf{Purifier}. The \textbf{Purifier} first adds noise to the adversarial example via the forward diffusion process and then runs the reverse process to obtain the purified example. Subsequently, the purified example is fed into the ASR system to get predictions. Without the \textbf{Purifier}, the adversarial example is fed into the ASR system directly. Inspired by \cite{wu2023defending}.}
    \label{fig: AudioPure}
\end{figure*}

Another approach for safeguarding ASR systems is detecting adversarial attacks. There are two main methods: one involves constructing a specialised classifier, while the other can generalise to a wider range of threats without needing a specialised classifier.
A method in the first category \cite{samizade2020adversarial} uses a convolutional neural network with small kernels to identify subtle perturbations in adversarial examples. This was enhanced in \cite{nielsen2023leveraging} by incorporating filter bank-based features for better detection. However, these methods are less effective against attacks dissimilar to those used in training. 
This motivates the second category of methods. 
These works either leverages dependencies among neighbouring sounds in audio sequences \cite{yang2018characterizing} or common signal processing transformations \cite{WaveGuard} for binary classification, 
achieving state-of-the-art results on strong adaptive attacks.
Another approach \cite{kwon2023audio} uses ASR classification scores to detect attacks without model training, but it fails when the classification scores are unknown. All these methods \cite{yang2018characterizing, WaveGuard, kwon2023audio} were tested on only 100 samples from the Mozilla Common Voice (MCV) dataset \cite{ardila2020common}.
We introduce a novel approach using DMs for detecting adversarial attacks. Our method compares ASR system outputs of non-purified and purified speech signals, tested on a substantially larger numbers of samples.




Our contributions are multifold. First, we explore using a pre-trained DDPM to defend against adversarial attacks on audio at the sentence level and analyse its behaviour. Next, we systematically study the impact of varying forward diffusion steps on ASR performance for both clean speech and adversarial examples. Extensive experiments on a subset of the MCV dataset \cite{ardila2020common} show that while the pre-trained DDPM can completely defend against adversarial attacks, ASR performance on clean speech degrades. We also found that two diffusion steps yield a 100\% defence success rate. Finally, we propose a novel purification-based detection method for adversarial attacks using the same pre-trained DDPM. Our method, tested on a larger subset of the MCV dataset than those used in the literature, can be run on consumer-grade hardware. Both our code and generated datasets are publicly available.\footnote{https://github.com/Kyhne/Detecting-and-Defending-Against-Adversarial-Attacks} The system architecture is illustrated in Fig.~\ref{fig: AudioPure}.

\section{Methods}

\subsection[Adversarial Attacking Method]{Adversarial Attacking Method}

The Carlini \& Wagner (C\&W) attacking method \cite{carlini2018audio} is chosen for its widespread adoption as one of the most successful targeted WB attacking methods for audio. It achieves \qty{100}{\percent} adversarial attack success rate on targeted attacks on a subset of the MCV dataset assuming the target is theoretically reachable \cite{carlini2018audio}.  
In the method, the added adversarial perturbation $\bm{\delta}$ is optimised by iteratively solving the optimisation problem
\begin{align}\label{eq:CWperturb}
    \underset{\bm{\delta}}{\text{minimise}} \ & \norm{ \bm{\delta} }_2^2+c\cdot \ell (\bm{x}+\bm{\delta},\bm{t}) \\
    &\text{such that}\ dB(\bm{\delta})-dB(\bm{x}) \leq\tau,
\end{align}
where $c$ is a regularisation term, $\ell(\cdot)$ is the connectionist temporal classification loss function \cite{CTC}, $\bm{x}\in\mathbb{R}^n$ is the original example, and $\bm{t}$ is the desired target phrase. 
Furthermore, 
the constant $\tau$ is initially sufficiently large to ensure a partial solution $\bm{\delta}^*$ exists, and then in each iteration $\tau$ is reduced until no solution can be found. 
\begin{figure}[ht]  
    \centering
    \input{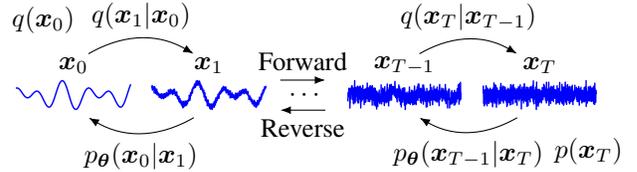}
    \caption{The forward and reverse diffusion process in DMs. 
    }
    \label{fig:DDPM}
\end{figure}

\subsection{Denoising Diffusion Probabilistic Models}

For DMs, this work uses the DDPM framework from \cite{ho2020denoising}. 
A DM consists of a forward diffusion process and a reverse diffusion process, defined by Markov chains, as shown in Fig.~\ref{fig:DDPM}.
The forward process gradually adds noise to the input data until the distribution of the noisy data approximately equals a standard Gaussian distribution. The reverse process is para\-meterised by a deep neural network that takes the approximately standard Gaussian noise as input and gradually denoises the data to recover clean data.

Formally, let $\bm{x}_0\in\mathbb{R}^n$ and $q(\bm{x}_0)$ be an unknown data distribution.
For each data point $\bm{x}_0\sim q(\bm{x}_0)$, a forward Markov chain is formed such that 
\begin{align}
q(\bm{x}_t\vert\bm{x}_{t-1}) = \mathcal{N}\big(\bm{x}_t\big\vert\sqrt{1-\beta_t}\bm{x}_{t-1},\,\beta_t\bm{I}\big), \ 1\leq t\leq T,
\end{align}
based on a pre-determined noise variance schedule $\{\beta_t\}_{t=1}^T$, where $0<\beta_t<1$.
Using the reparameterisation trick \cite{varautoencoder} with $\alpha_t = 1-\beta_t$ and $\overline{\alpha}_t = \prod_{i=0}^t\alpha_i$ results in 
\begin{align}
q(\bm{x}_t|\bm{x}_0) = \mathcal{N}\big(\bm{x}_t\big\vert\sqrt{\overline{\alpha}_t}\bm{x}_0,(1-\overline{\alpha}_t)\bm{I}\big).
\end{align}
The sequence $\{\beta_t\}_{t=1}^T$ is chosen such that $\overline{\alpha}_T\approx 0$, which results in $q(\bm{x}_T|\bm{x}_0)\approx \mathcal{N}(\bm{0}, \bm{I})$, and thus $q(\bm{x}_T)\approx \mathcal{N}(\bm{x}_T|\bm{0},\bm{I})$. 
Since the reverse diffusion process is intractable, the denoising process $q(\bm{x}_{t-1}|\bm{x}_t)$ is approximated by a learnable Markov chain with parameters $\bm{\theta}$ defined by
\begin{align}
    p_{\bm{\theta}}(\bm{x}_{t-1}\vert\bm{x}_t) = \mathcal{N}\big(\bm{x}_{t-1}\vert\bm{\mu}_{\bm{\theta}}(\bm{x}_t,t),\, \bm{\Sigma}_{\bm{\theta}}(\bm{x}_t, t)\big)
\end{align}
given a prior distribution  $p(\bm{x}_T)=\mathcal{N}(\bm{x}_T\vert\bm{0},\bm{I})$. Specifically,
\begin{align}
     \bm{\mu}_{\bm{\theta}}(\bm{x}_t,t) &= \frac{1}{\sqrt{\alpha_t}}\left(\bm{x}_t - \frac{1-\alpha_t}{\sqrt{1-\overline{\alpha}_t}}\hat{\bm{\epsilon}}_{\bm{\theta}}(\bm{x}_t,t)\right), \\\bm{\Sigma}_{\bm{\theta}}(\bm{x}_t,t)&= \frac{1-\overline{\alpha}_{t-1}}{1-\overline{\alpha}_t}\beta_t \bm{I},\label{eq:Niko}
\end{align}
where $\hat{\bm{\epsilon}}_{\bm{\theta}}:\mathbb{R}^n\times \mathbb{N}\to \mathbb{R}^n$ is a deep neural network predicting the noise $\bm{\epsilon}_0\sim\mathcal{N}(\bm{\epsilon}|\bm{0},\bm{I})$ in the forward process. 
The choice of $\bm{\Sigma}_{\bm{\theta}}(\bm{x}_t,t)$ is based on \cite{wu2023defending} as it delivers superior results for adversarial purification on audio.

In this work, we use an adversarial purification-based method for defence denoted as \textbf{Purifier}, depicted in Fig.~\ref{fig: AudioPure}. The \textbf{Purifier} is based on the plug-and-play method \audiopure \cite{wu2023defending}, where we omit the Wave2Mel module that extracts Mel spectrograms as we focus solely on waveform ASR systems.

\subsection{Method for Defending Against Adversarial Attacks}
For defence, a waveform input is passed through the \textbf{Purifier} and a pre-trained ASR system, sequentially.
The \textbf{Purifier} is based on a pre-trained DDPM: Diff\-Wave \cite{kong2021diffwave}. 
DiffWave is a $36$-layer, $6.91$M parameter residual neural network \cite{he2016deep} and uses a bidirectional dilated convolution architecture. The kernel size is $3$, the dilation cycle is $[1,2,4,\dots,2048]$, and the number of residual channels is $C=256$. 
Furthermore, the number of forward steps is $T=200$, and the noise schedule is linearly spaced for $0.0001\leq \beta_t \le 0.02$.

Let $\bm{x}_{\mathrm{adv}} \in \mathbb{R}^n$ be the adversarial waveform input.
To override the adversarial perturbations,
the \textbf{Purifier} adds noise to $\bm{x}_{\mathrm{adv}}$ via the forward diffusion process. 
From the $T$ forward diffusion steps, only the first $n^{\star}\in\{1,2,\dots,T\}$ steps are used. This prevents excessive degradation of the original waveform, which could hinder recovery and lead to misclassifications. The reverse diffusion process then aims to reconstruct the clean signal from the noisy input. The grey box in Fig.~\ref{fig: AudioPure} depicts this process.
Formally, the \textbf{Purifier} takes $\bm{x}_{\mathrm{adv}}$ and $n^{\star}$ as inputs, and returns the purified-speech signal leading to the function $\textbf{Purifier}: \mathbb{R}^n\times\mathbb{N}\to\mathbb{R}^n$. Next, the ASR system $F:\mathbb{R}^n\to\mathbb{R}^d$ is applied to the purified-speech signal, where $d$ is the length of the output. Finally, the \textbf{Purifier} and the ASR system are combined into a defended speech system $\textbf{SS}:\mathbb{R}^n\times\mathbb{N}\to\mathbb{R}^d$ shown in Fig.~\ref{fig: AudioPure} and given by
\begin{align}\textbf{SS}\left(\bm{x}_{\mathrm{adv}},n^{\star}\right) = F\left(\textbf{Purifier}\left(\bm{x}_{\mathrm{adv}},n^{\star}\right)\right)\!.
\end{align}

\subsection{Method for Detecting Adversarial Attacks}
We propose a detection method using a binary classifier that labels inputs as either adversarial or benign. To classify an input sentence $\bm{x}_{\mathrm{in}}$, the character error rate (CER) is calculated between the non-purified speech ASR output $F(\bm{x}_{\mathrm{in}})$ and the purified speech ASR output $\mathbf{SS}(\bm{x}_{\mathrm{in}},n^\star)$.
If $\mathrm{CER}(\mathbf{SS}(\bm{x}_{\mathrm{in}},n^\star),F(\bm{x}_{\mathrm{in}}))>\Omega$ for a pre-determined threshold $\Omega$ and forward diffusion steps $n^\star$, $\bm{x}_{\mathrm{in}}$ is classified as adversarial,
otherwise as benign.
The threshold $\Omega$ and forward diffusion steps $n^{\star}$ are found through a grid search systematically exploring different values of hyperparameters.
CER is used as opposed to word error rate (WER), since preliminary experiments showed that CER yields better results.

\section{Experiments and Results}

\subsection{Defending Against Adversarial Attacks}\label{subsec:MethodA}
We experiment with adversarial defence and purification for sentences with varying forward diffusion steps.
In all experiments $n^{\star} \in \{1,2,3,4,5\}$. The dataset and models were selected due to the high adversarial attack success rate on the ASR system and dataset\cite{carlini2018audio}.

\smallskip

\noindent\textbf{Datasets.} 
From the MCV dataset \cite{ardila2020common}, $300$ short (S), $300$ medium (M), and $300$ long (L) English sentences of length $1$-$2$, $3$-$4$, and $6$-$7$ seconds, respectively, are chosen. Only audio files containing more than \qty{68}{\percent} speech have been chosen using the open-source robust Voice Activity Detection (rVAD) algorithm \cite{tan2020rvad}. All sentences are attacked with the same WB attack using the same target for sentences of the same length:
\begin{itemize}
    \item \textbf{Short target:} \textit{open all doors}. 
    \item \textbf{Medium target:} \textit{switch off internet connection}. 
    \item \textbf{Long target:} \textit{i need a reservation for sixteen people at the seafood restaurant down the street}.
\end{itemize}

\smallskip
\noindent\textbf{Attack Method.} The C\&W \cite{carlini2018audio} method is used, where the maximum amount of iterations is $5000$ and the learning rate is $10$ as in \cite{carlini2018audio}.

\smallskip
\noindent\textbf{Models.} The unconditional version of DiffWave \cite{kong2021diffwave} with the officially provided pre-trained checkpoint is utilised as the defensive waveform purifier. 
A pre-trained end-to-end Deep Speech system\footnote{
https://github.com/mozilla/DeepSpeech/releases/tag/v0.9.3, 2020} \cite{DeepSpeechNR1} is used as an ASR system.

\smallskip
\noindent \textbf{Evaluation metrics.} 
The adversarial attack success rate is computed as the percentage of times the ASR system transcribes exactly the target of the attack.

The ASR performance is represented as $1-\mathrm{CER}$, where CER is calculated between the transcribed output and the ground truth label.
CER is used instead of WER, since error rate at the character level provides higher granularity.

\subsection{Detecting Adversarial Attacks}

In terms of \textbf{datasets, attack method, and models}, adversarial detection is performed on sentences using the same configurations as detailed in Section~\ref{subsec:MethodA}.

Attacking all $900$ benign sentences yields $900$ adversarial sentences. As such, we generate $300$ benign and $300$ adversarial sentences for each sentence lengths resulting in a dataset containing 1800 sentences.
For each sentence length, we split the data into two parts: the first \qty{10}{\percent} for determining the threshold $\Omega$ and the number of forward diffusion steps $n^{\star}$, and the next \qty{90}{\percent} for testing, each part containing \qty{50}{\percent} adversarial and \qty{50}{\percent} benign sentences. 
For each $n^{\star}\in\{1,2,3,4,5\}$, a grid search is performed to find the $\Omega$ that maximises the adversarial detection accuracy across all sentence lengths, leading to the global hyperparameters $(n^{\star},\Omega)$.
\aboverulesep=0ex
\belowrulesep=0ex
\setlength\tabcolsep{1.9em}
\begin{table}[h]
    \centering
    \caption{Average runtime per sentence length in seconds averaged over $300$ examples.}
    \begin{tabular}{@{}lc|ccc@{}}
    \toprule
         & $n^\star$ & S & M & L\\\hline
        \textbf{Purifier} & $1$ & $0.51$  & $0.88$  & $1.67$ \\
        \textbf{Purifier} & $5$ & $1.44$  & $2.68$  & $4.84$ \\
        Deep Speech & $-$ & $0.68$  & $1.30$  & $2.22$ \\
        Detection & $2$ & $2.17$ & $4.19$ & $8.40$\\
    \bottomrule
    \end{tabular}
    \label{tab:runtime}
\end{table}

\setlength\tabcolsep{1.65em}
\begin{table*}[t]
    \centering
    \caption{ASR performance measured at character level. 
    }
    \begin{tabular}{@{}l|ccccccc@{}}
    \toprule
    $n^{\star}$  & Clean-S         & WB-S                & Clean-M         & WB-M          & Clean-L          & WB-L          \\ \hline
    $0$ & $\bm{78.00}$   & $19.41$             & $\bm{86.01}$   & $20.58$       & $\bm{86.07}$    & $24.67$       \\
    $1$  & $72.10\pm0.66$ & $42.62\pm0.34$      & $75.71\pm0.45$ & $34.56\pm0.15$ & $73.99\pm0.10$  & $35.58\pm0.08$ \\
    $2$ & $65.32\pm0.73$ & $45.79\pm0.74$      & $67.28\pm0.47$ & $39.74\pm0.13$       & $64.84\pm0.46$  & $40.05\pm0.17$       \\
    $3$  & $58.88\pm0.28$ & $47.51\pm0.36$ & $60.98\pm0.26$ & $43.97\pm0.66$       & $57.86\pm0.22$  & $43.94\pm0.13$       \\
    $4$  & $54.87\pm0.62$ & $\bm{47.64\pm0.58}$      & $55.76\pm0.30$ & $\bm{46.15\pm0.15}$  & $52.16\pm0.21$  & $\bm{44.90\pm0.20}$  \\
    $5$ & $50.20\pm1.75$ & $46.41\pm0.98$             & $50.96\pm0.39$ & $46.02\pm0.27$        & $47.46\pm0.21$ & $44.39\pm0.16$       \\
    \bottomrule

\end{tabular}
    \label{tab:classified}
\end{table*}

\renewcommand{\arraystretch}{1.3}
\setlength{\tabcolsep}{1.075em}
\begin{table*}[b]
    \renewcommand\thetable{IV}
    \centering
    \caption{Classification scores from the detection experiment given $\Omega=0.57$ and $n^{\star}=2$ for the WB attacks. TN stands for true negative, FN for false negative, FP for false positive, and TP for true positive.}
    
    \begin{tabular}{@{}l|c|c|c@{}}
    \toprule
         Length & S & M & L \\\hline
         &&&\\[-4mm]
        \begin{tabular}{c|c}
           TN  & FP \\\hline
           FN  & TP
        \end{tabular}        &      \begin{tabular}{c|c}
                                            $0.76\pm0.00$      & $0.24\pm0.00$      \\\hline
                                            $\bm{0.02}\pm0.00$ & $\bm{0.98}\pm0.00$
                                        \end{tabular}           &   \begin{tabular}{c|c}
                                                                        $\bm{0.81}\pm0.00$ & $\bm{0.19}\pm0.00$ \\\hline
                                                                        $0.06\pm0.00$      & $0.94\pm0.00$
                                                                    \end{tabular}            &   \begin{tabular}{c|c}
                                                                                                    $0.77\pm0.00$ & $0.23\pm0.00$ \\\hline
                                                                                                    $0.04\pm0.00$ & $0.96\pm0.00$
                                                                                                \end{tabular}        \\
        &&&\\[-4mm]\hline
        Accuracy     & $0.87\pm0.00$       & $\bm{0.88}\pm0.00$ & $0.87\pm0.00$\\ \hline
        AUROC     & $\bm{0.92}\pm0.00$       & $0.90\pm0.00$ & $0.89\pm0.00$\\
    \bottomrule
    \end{tabular}
    \label{tab:detection}
\end{table*}

\renewcommand{\arraystretch}{1}
\setlength\tabcolsep{1.7em}
\begin{table}[t]
    \renewcommand\thetable{III}
    \centering
    \caption{Adversarial attack success rates.}
    \begin{tabular}{@{}l|ccc@{}}
    \toprule
    $n^{\star}$                  & WB-S               & WB-M               & WB-L               \\\hline
    $0$ &   $71.33$            & $90.67$            & $94.33$            \\
    $1$ & $1.27\pm 0.09$     & $0.33\pm 0.12$ & $0.03\pm 0.01$      \\
    $2$ & $\bm{0.00\pm0.00}$ & $\bm{0.00\pm 0.00}$      & $\bm{0.00\pm 0.00}$ \\
    $3$ & $0.00\pm 0.00$     & $0.00\pm 0.00$      & $0.00\pm 0.00$      \\
    $4$ & $0.00\pm 0.00$     & $0.00\pm 0.00$      & $0.00\pm 0.00$      \\
    $5$ & $0.00\pm 0.00$      & $0.00\pm 0.00$      & $0.00\pm 0.00$      \\
    \bottomrule
\end{tabular}
    \label{tab:target} 
\end{table}

\smallskip

\noindent \textbf{Evaluation metrics.} The detection method is evaluated using a confusion matrix as well as the area under the receiver operating characteristic curve (AUROC) score. In the confusion matrix, a positive or negative outcome refers to an input being classified as adversarial or benign, respectively. 

\subsection{Hardware and Runtime}
All experiments are performed on an NVIDIA RTX $3060$ GPU with AMD Ryzen $7$ $5800$H @ $\qty{3.20}{GHz}$ and $\qty{16}{GB}$ RAM. The runtime for the experiment can be seen in Table~\ref{tab:runtime}. The reported times in Table~\ref{tab:runtime} show that the proposed method might be used with constrained computational resources.

\subsection{Results}
In Tables \ref{tab:classified}, \ref{tab:target} and \ref{tab:detection}, the \qty{95}{\percent} confidence intervals are reported over $10$ runs, and $n^{\star}=0$ indicates that the \textbf{Purifier} has not been applied.

\smallskip
\noindent \textbf{Adversarial Purification.}
Comparing Clean-(S,M,L) with WB-(S,M,L) in Table~\ref{tab:classified} (row $n^{\star}=0$) shows that the WB attack decreases relative ASR performance on clean speech by at least \qty{71}{\percent} for all sentence lengths.
ASR performance on adversarial examples (or robustness accuracy) is the highest when $n^\star=4$ for WB-(S,M,L). 
As seen in Table~\ref{tab:target}, we do not achieve \qty{100}{\percent} adversarial attack success rate as in \cite{carlini2018audio}, since we did not account for the theoretical reachability of the targets.
Table~\ref{tab:target} further shows that WB attacks are completely defended against when $n^\star\geq2$ for all sentence lengths.

\smallskip
\noindent \textbf{Adversarial detection.}
Based on the grid search, the optimal thres\-hold was found to be $\Omega=0.57$ with $n^{\star}=2$ for all sentence lengths. 
Using these hyperparameters, the accuracy is at least \qty{87}{\percent}, the true positive rate is at least \qty{94}{\percent}, and the AUROC score is at least 0.89 as shown in Table~\ref{tab:detection}. Choosing $0.55 \leq \Omega \leq 0.59$ yields accuracies and AUROC scores varying approximately $\pm\qty{1.00}{pp}$, indicating that the detection algorithm is resilient to the threshold change.

\section{Discussion}

Tables~\ref{tab:classified} and \ref{tab:target} show
that the \textbf{Purifier} can be used to purify adversarial examples and completely defend against WB attacks on sentences. 
However, the clean signals get misclassified more often when $n^\star$ increases, due to distortion.

Table~\ref{tab:detection} shows that the proposed novel purification-based adversarial detection method achieves high accuracies and true positive rates, which
 is desirable as we aim to ensure no adversarial examples go undetected.

Methods relying on training a dedicated detection classifier, e.g., \cite{samizade2020adversarial}, achieve high detection accuracies for seen attacks, while their performance degrades dramatically for unseen examples. 
Our training-free approach 
should not be limited to specific adversarial attacks.
By comparing our approach to the training-free methods in \cite{yang2018characterizing, kwon2023audio}, we achieve similar AUROC scores and accuracies with a much larger dataset.

\section{Conclusion}
\noindent
In this paper, we leveraged a pre-trained DM to defend against adversarial attacks. 
Comprehensive experiments indicate that increasing the number of forward diffusion steps in the diffusion process improves ASR performance on adversarial examples at the cost of clean speech ASR performance. Two forward diffusion steps ensure an adversarial attack success rate of \qty{0.00}{\percent}.
Finally, we introduced a novel approach utilising pre-trained DMs for detecting unknown adversarial attacks on sentences. Experiments have shown its effectiveness, achieving high AUROC scores, true positive rates, and accuracies. 

\bibliographystyle{IEEEtran}
\bibliography{litteratur}

\begin{thebibliography}{10}
\providecommand{\url}[1]{#1}
\csname url@samestyle\endcsname
\providecommand{\newblock}{\relax}
\providecommand{\bibinfo}[2]{#2}
\providecommand{\BIBentrySTDinterwordspacing}{\spaceskip=0pt\relax}
\providecommand{\BIBentryALTinterwordstretchfactor}{4}
\providecommand{\BIBentryALTinterwordspacing}{\spaceskip=\fontdimen2\font plus
\BIBentryALTinterwordstretchfactor\fontdimen3\font minus \fontdimen4\font\relax}
\providecommand{\BIBforeignlanguage}[2]{{%
\expandafter\ifx\csname l@#1\endcsname\relax
\typeout{** WARNING: IEEEtran.bst: No hyphenation pattern has been}%
\typeout{** loaded for the language `#1'. Using the pattern for}%
\typeout{** the default language instead.}%
\else
\language=\csname l@#1\endcsname
\fi
#2}}
\providecommand{\BIBdecl}{\relax}
\BIBdecl

\bibitem{DBLP:journals/corr/SzegedyZSBEGF13}
C.~Szegedy, W.~Zaremba, I.~Sutskever, J.~Bruna, D.~Erhan, I.~J. Goodfellow, and R.~Fergus, ``Intriguing properties of neural networks,'' in \emph{2nd International Conference on Learning Representations}, 2014.

\bibitem{43405}
I.~Goodfellow, J.~Shlens, and C.~Szegedy, ``Explaining and harnessing adversarial examples,'' in \emph{International Conference on Learning Representations}, 2015.

\bibitem{papernot2016limitations}
N.~Papernot, P.~McDaniel, S.~Jha, M.~Fredrikson, Z.~B. Celik, and A.~Swami, ``The limitations of deep learning in adversarial settings,'' in \emph{2016 IEEE European symposium on security and privacy (EuroS\&P)}.\hskip 1em plus 0.5em minus 0.4em\relax IEEE, 2016, pp. 372--387.

\bibitem{cisse2017houdini}
M.~M. Cisse, Y.~Adi, N.~Neverova, and J.~Keshet, ``Houdini: Fooling deep structured visual and speech recognition models with adversarial examples,'' \emph{Advances in neural information processing systems}, vol.~30, 2017.

\bibitem{yuan2018commandersong}
X.~Yuan, Y.~Chen, Y.~Zhao, Y.~Long, X.~Liu, K.~Chen, S.~Zhang, H.~Huang, X.~Wang, and C.~A. Gunter, ``$\{$CommanderSong$\}$: a systematic approach for practical adversarial voice recognition,'' in \emph{27th USENIX security symposium (USENIX security 18)}, 2018, pp. 49--64.

\bibitem{qin2019imperceptible}
Y.~Qin, N.~Carlini, G.~Cottrell, I.~Goodfellow, and C.~Raffel, ``Imperceptible, robust, and targeted adversarial examples for automatic speech recognition,'' in \emph{International conference on machine learning}.\hskip 1em plus 0.5em minus 0.4em\relax PMLR, 2019, pp. 5231--5240.

\bibitem{taori2019targeted}
R.~Taori, A.~Kamsetty, B.~Chu, and N.~Vemuri, ``Targeted adversarial examples for black box audio systems,'' in \emph{2019 IEEE security and privacy workshops (SPW)}.\hskip 1em plus 0.5em minus 0.4em\relax IEEE, 2019, pp. 15--20.

\bibitem{du2020sirenattack}
T.~Du, S.~Ji, J.~Li, Q.~Gu, T.~Wang, and R.~Beyah, ``Sirenattack: Generating adversarial audio for end-to-end acoustic systems,'' in \emph{Proceedings of the 15th ACM Asia conference on computer and communications security}, 2020, pp. 357--369.

\bibitem{abdullah2021hear}
H.~Abdullah, M.~S. Rahman, W.~Garcia, K.~Warren, A.~S. Yadav, T.~Shrimpton, and P.~Traynor, ``Hear" no evil", see" kenansville": Efficient and transferable black-box attacks on speech recognition and voice identification systems,'' in \emph{2021 IEEE Symposium on Security and Privacy (SP)}.\hskip 1em plus 0.5em minus 0.4em\relax IEEE, 2021, pp. 712--729.

\bibitem{chen2021real}
G.~Chen, S.~Chenb, L.~Fan, X.~Du, Z.~Zhao, F.~Song, and Y.~Liu, ``Who is real bob? adversarial attacks on speaker recognition systems,'' in \emph{2021 IEEE Symposium on Security and Privacy (SP)}.\hskip 1em plus 0.5em minus 0.4em\relax IEEE, 2021, pp. 694--711.

\bibitem{herff2016automatic}
C.~Herff and T.~Schultz, ``Automatic speech recognition from neural signals: a focused review,'' \emph{Frontiers in neuroscience}, vol.~10, p. 429, 2016.

\bibitem{yang2018characterizing}
Z.~Yang, B.~Li, P.-Y. Chen, and D.~Song, ``Characterizing audio adversarial examples using temporal dependency,'' \emph{International Conference on Learning Representations}, 2019.

\bibitem{rajaratnam2018speech}
K.~Rajaratnam, B.~Alshemali, and J.~Kalita, ``Speech coding and audio preprocessing for mitigating and detecting audio adversarial examples on automatic speech recognition,'' 2018.

\bibitem{distillation2016}
N.~Papernot, P.~McDaniel, X.~Wu, S.~Jha, and A.~Swami, ``Distillation as a defense to adversarial perturbations against deep neural networks,'' in \emph{2016 IEEE symposium on security and privacy (SP)}.\hskip 1em plus 0.5em minus 0.4em\relax IEEE, 2016, pp. 582--597.

\bibitem{chen2022towards}
G.~Chen, Z.~Zhao, F.~Song, S.~Chen, L.~Fan, F.~Wang, and J.~Wang, ``Towards understanding and mitigating audio adversarial examples for speaker recognition,'' \emph{IEEE Transactions on Dependable and Secure Computing}, vol.~20, no.~5, pp. 3970--3987, 2022.

\bibitem{carlini2018audio}
N.~Carlini and D.~Wagner, ``Audio adversarial examples: Targeted attacks on speech-to-text,'' in \emph{2018 IEEE Security and Privacy Workshops (SPW)}.\hskip 1em plus 0.5em minus 0.4em\relax IEEE, 2018, pp. 1--7.

\bibitem{carlini2017towards}
------, ``Towards evaluating the robustness of neural networks,'' in \emph{2017 IEEE symposium on security and privacy (sp)}.\hskip 1em plus 0.5em minus 0.4em\relax IEEE, 2017, pp. 39--57.

\bibitem{madry2017towards}
A.~Madry, A.~Makelov, L.~Schmidt, D.~Tsipras, and A.~Vladu, ``Towards deep learning models resistant to adversarial attacks,'' \emph{Proceedings of the International Conference on Representation Learning}, 2017.

\bibitem{wu2023defending}
S.~Wu, J.~Wang, W.~Ping, W.~Nie, and C.~Xiao, ``Defending against adversarial audio via diffusion model,'' \emph{The Eleventh International Conference on Learning Representations}, 2023.

\bibitem{sohl2015deep}
J.~Sohl-Dickstein, E.~Weiss, N.~Maheswaranathan, and S.~Ganguli, ``Deep unsupervised learning using nonequilibrium thermodynamics,'' in \emph{International conference on machine learning}.\hskip 1em plus 0.5em minus 0.4em\relax PMLR, 2015, pp. 2256--2265.

\bibitem{ho2020denoising}
J.~Ho, A.~Jain, and P.~Abbeel, ``Denoising diffusion probabilistic models,'' \emph{Advances in Neural Information Processing Systems}, vol.~33, pp. 6840--6851, 2020.

\bibitem{nichol2021improved}
A.~Q. Nichol and P.~Dhariwal, ``Improved denoising diffusion probabilistic models,'' in \emph{International Conference on Machine Learning}.\hskip 1em plus 0.5em minus 0.4em\relax PMLR, 2021, pp. 8162--8171.

\bibitem{kong2021diffwave}
Z.~Kong, W.~Ping, J.~Huang, K.~Zhao, and B.~Catanzaro, ``Diffwave: A versatile diffusion model for audio synthesis,'' \emph{ICLR 2021 (oral)}, 2021.

\bibitem{samizade2020adversarial}
S.~Samizade, Z.-H. Tan, C.~Shen, and X.~Guan, ``Adversarial example detection by classification for deep speech recognition,'' in \emph{ICASSP 2020-2020 IEEE International Conference on Acoustics, Speech and Signal Processing (ICASSP)}.\hskip 1em plus 0.5em minus 0.4em\relax IEEE, 2020, pp. 3102--3106.

\bibitem{nielsen2023leveraging}
C.~H. Nielsen and Z.-H. Tan, ``Leveraging domain features for detecting adversarial attacks against deep speech recognition in noise,'' \emph{IEEE Open Journal of Signal Processing}, vol.~4, pp. 179--187, 2023.

\bibitem{WaveGuard}
S.~Hussain, P.~Neekhara, S.~Dubnov, J.~McAuley, and F.~Koushanfar, ``{WaveGuard}: Understanding and mitigating audio adversarial examples,'' in \emph{30th USENIX Security Symposium (USENIX Security 21)}.\hskip 1em plus 0.5em minus 0.4em\relax USENIX Association, Aug. 2021, pp. 2273--2290.

\bibitem{kwon2023audio}
H.~Kwon and S.-H. Nam, ``Audio adversarial detection through classification score on speech recognition systems,'' \emph{Computers \& Security}, vol. 126, p. 103061, 2023.

\bibitem{ardila2020common}
R.~Ardila, M.~Branson, K.~Davis, M.~Henretty, M.~Kohler, J.~Meyer, R.~Morais, L.~Saunders, F.~M. Tyers, and G.~Weber, ``Common voice: A massively-multilingual speech corpus,'' in \emph{Proceedings of the 12th Conference on Language Resources and Evaluation (LREC 2020)}, 2020, pp. 4211--4215.

\bibitem{CTC}
A.~Graves, S.~Fern{\'a}ndez, F.~Gomez, and J.~Schmidhuber, ``Connectionist temporal classification: labelling unsegmented sequence data with recurrent neural networks,'' in \emph{Proceedings of the 23rd International Conference on Machine learning}, 2006, pp. 369--376.

\bibitem{varautoencoder}
D.~P. Kingma and M.~Welling, ``Auto-encoding variational bayes,'' in \emph{International Conference on Learning Representations}, 2014.

\bibitem{he2016deep}
K.~He, X.~Zhang, S.~Ren, and J.~Sun, ``Deep residual learning for image recognition,'' in \emph{Proceedings of the IEEE conference on computer vision and pattern recognition}, 2016, pp. 770--778.

\bibitem{tan2020rvad}
Z.-H. Tan, A.~kr. Sarkar, and N.~Dehak, ``r{VAD}: An unsupervised segment-based robust voice activity detection method,'' \emph{Computer speech \& language}, vol.~59, pp. 1--21, 2020, \url{https://github.com/zhenghuatan/rVAD}.

\bibitem{DeepSpeechNR1}
A.~Hannun, C.~Case, J.~Casper, B.~Catanzaro, G.~Diamos, E.~Elsen, R.~Prenger, S.~Satheesh, S.~Sengupta, A.~Coates \emph{et~al.}, ``Deep speech: Scaling up end-to-end speech recognition,'' \emph{arXiv preprint arXiv:1412.5567}, 2014.

\end{thebibliography}


\begin{thebibliography}{00}
\bibitem{b1} G. Eason, B. Noble, and I. N. Sneddon, ``On certain integrals of Lipschitz-Hankel type involving products of Bessel functions,'' Phil. Trans. Roy. Soc. London, vol. A247, pp. 529--551, April 1955.
\bibitem{b2} J. Clerk Maxwell, A Treatise on Electricity and Magnetism, 3rd ed., vol. 2. Oxford: Clarendon, 1892, pp.68--73.
\bibitem{b3} I. S. Jacobs and C. P. Bean, ``Fine particles, thin films and exchange anisotropy,'' in Magnetism, vol. III, G. T. Rado and H. Suhl, Eds. New York: Academic, 1963, pp. 271--350.
\bibitem{b4} K. Elissa, ``Title of paper if known,'' unpublished.
\bibitem{b5} R. Nicole, ``Title of paper with only first word capitalized,'' J. Name Stand. Abbrev., in press.
\bibitem{b6} Y. Yorozu, M. Hirano, K. Oka, and Y. Tagawa, ``Electron spectroscopy studies on magneto-optical media and plastic substrate interface,'' IEEE Transl. J. Magn. Japan, vol. 2, pp. 740--741, August 1987 [Digests 9th Annual Conf. Magnetics Japan, p. 301, 1982].
\bibitem{b7} M. Young, The Technical Writer's Handbook. Mill Valley, CA: University Science, 1989.
\bibitem{b8} D. P. Kingma and M. Welling, ``Auto-encoding variational Bayes,'' 2013, arXiv:1312.6114. [Online]. Available: https://arxiv.org/abs/1312.6114
\bibitem{b9} S. Liu, ``Wi-Fi Energy Detection Testbed (12MTC),'' 2023, gitHub repository. [Online]. Available: https://github.com/liustone99/Wi-Fi-Energy-Detection-Testbed-12MTC
\bibitem{b10} ``Treatment episode data set: discharges (TEDS-D): concatenated, 2006 to 2009.'' U.S. Department of Health and Human Services, Substance Abuse and Mental Health Services Administration, Office of Applied Studies, August, 2013, DOI:10.3886/ICPSR30122.v2
\bibitem{b11} K. Eves and J. Valasek, ``Adaptive control for singularly perturbed systems examples,'' Code Ocean, Aug. 2023. [Online]. Available: https://codeocean.com/capsule/4989235/tree
\end{thebibliography}

\end{document}